\documentclass[sigconf]{acmart}
\renewcommand\footnotetextcopyrightpermission[1]{} 
\usepackage{booktabs} 
\usepackage{graphicx}
\usepackage{subcaption}
\usepackage{dblfloatfix}
\usepackage{url}
\usepackage{makecell}
\usepackage{multirow}
\usepackage{enumitem}

\usepackage[group-separator={,}]{siunitx}
\begin{document}
\title{A Ground-Truth Data Set and a Classification Algorithm for Eye Movements in 360-degree Videos}
\renewcommand{\shorttitle}{A Ground-Truth Data Set and a Classification Algorithm for Eye Movements in 360-degree Videos} 

\author{Ioannis Agtzidis}
\affiliation{%
  \institution{Technical University of Munich}
  \streetaddress{Arcisstr. 21}
  \city{Munich} 
  \country{Germany} 
  \postcode{80333}
}
\email{ioannis.agtzidis@tum.de}

\author{Mikhail Startsev}
\affiliation{%
  \institution{Technical University of Munich}
  \streetaddress{Arcisstr. 21}
  \city{Munich} 
  \country{Germany} 
  \postcode{80333}
}
\email{mikhail.startsev@tum.de}

\author{Michael Dorr}
\affiliation{%
  \institution{Technical University of Munich}
  \streetaddress{Arcisstr. 21}
  \city{Munich} 
  \country{Germany} 
  \postcode{80333}
}
\email{michael.dorr@tum.de}

\renewcommand{\shortauthors}{I. Agtzidis et al.}

\begin{abstract}

The segmentation of a gaze trace into its constituent eye movements has been
actively researched since the early days of eye tracking. As we move towards
more naturalistic viewing conditions, the segmentation becomes even more challenging and
convoluted as more complex patterns emerge. The definitions and the  well-established 
methods that were developed for monitor-based eye tracking experiments 
are often not directly applicable to unrestrained
set-ups such as eye tracking in wearable contexts or with head-mounted displays. 
The main contributions of this work to the eye movement research for {360$^\circ$} content
are threefold: 
First, we collect, partially annotate, and make publicly available 
a new eye tracking data set, which consists of 13 participants viewing 15 video clips that are 
recorded in {360$^\circ$}.
Second, we propose a new two-stage pipeline for ground truth annotation 
of the traditional fixations, saccades, smooth pursuits, as well 
as (optokinetic) nystagmus, vestibulo-ocular reflex, and pursuit of moving objects performed 
exclusively via the movement of the head. 
A flexible user interface for this pipeline is implemented and 
made freely accessible for
use or modification.
Lastly, we develop and test a simple proof-of-concept algorithm for automatic
classification of all the eye movement types in our data set based on their
operational definitions that were used for manual annotation. The data set and 
the source code for both the annotation tool and the algorithm are publicly available
at \url{https://web.gin.g-node.org/ioannis.agtzidis/360_em_dataset}.

\end{abstract}

\maketitle

\section{Introduction}

Eye tracking offers a non-invasive insight into the underpinnings of the human visual
system, its mechanisms of perception and processing. 
The holy grail of eye tracking would be to enable pervasive accurate monitoring 
of gaze direction
and targets, as well as the situational context, in everyday life. While this is still 
unrealistic, there are several ways of approaching it already. 
One is creating lightweight eye tracking glasses, and several models have recently become commercially
available\footnote{\url{https://www.tobiipro.com/product-listing/tobii-pro-glasses-2/}}\footnote{\url{https://pupil-labs.com/pupil/}}\footnote{\url{https://www.ergoneers.com/en/hardware/dikablis-glasses/}}.
These also capture the potential gaze targets with a scene camera, but
dealing with the real world is difficult:
The camera recordings will be affected
by motion blur from head movements and locomotion, direct sunlight, etc. Also, 
only part of the full context can be captured by the scene camera.

Another approach to carrying out eye tracking experiments that are close to the real-world studies 
is utilising
head-mounted displays (HMDs) to substitute (or augment) reality with content of 
varying degree of controlledness: From simplified programmatically generated stimuli, 
to interactive rich environments, or full-360$^\circ$ recordings of the real world.
Immersion of virtual reality (VR) and HMD content depends on a variety of 
factors \cite{jennett2008measuring, cummings2016immersive} and is not fully understood yet,
but some realism is likely sacrificed in order to gain a higher
degree of control and relative ease of analysis.

Here, we wanted to precisely characterise eye movement behaviour in a scenario
that is as unconstrained 
as possible. We therefore presented 360$^\circ$ 
videos in an HMD with integrated eye tracking, 
which enabled us to approach
the naturally occurring complexity of reality, while maintaining some degree of control over 
the audiovisual content. Such videos are naturally 
affected by lighting artefacts and camera motion, but they
can be quality-checked before presentation, and head motion-induced 
blur is not present.

In any eye tracking set-up, however, there persists a major challenge -- accurately and
robustly classifying the eye movements. In the case when the participant's head motion is
restricted (e.g.\ by a chin bar), the types of eye movements are mostly well understood,
though not as well as we would like to believe \cite{hessels2018eye}, since researchers 
often disagree on fundamental definitions for fixations and saccades. Further complications arise 
when smooth pursuit (SP) is introduced into the mix because it is often thought of as 
a ``fixation on a moving target'' even though it can reach high speeds~\cite{meyer1985upper}.
Detecting SP algorithmically is notably a more challenging endeavour than detecting fixations and saccades,
even for state-of-the-art algorithms \cite{larsson2016smooth, startsev2018cnn}.
The distinction between various eye movements becomes even more complex
when the head of the recorded participant can move freely. There are no 
commonly used definitions in this area of research, and \cite{hessels2018eye} urge the 
community to make their eye movement definitions explicit, since currently 
the algorithmic approaches for ``fixation'' detection in VR
or mobile eye tracking imply very different underlying interpretations of this eye movement type, and 
comparing studies with different definitions is impossible or, at the very least,
confusing.
In VR, for example, the TobiiPro
software\footnote{\url{https://www.tobiipro.com/product-listing/vr-analytics/}}
is using dwells on the same VR object (by intersecting the gaze ray with the virtual scene) as a substitute for 
fixations (very similar to I-AOI \cite{salvucci2000identifying}), 
while their own mobile and even 360$^\circ$
video-based\footnote{\url{https://www.tobiipro.com/product-listing/tobii-pro-lab/}} 
eye tracking solutions use simple speed thresholding
\cite{olsen2012tobii}. This means that 
the implied fixation definitions in the two cases differ wildly: The first would replace 
all pursuits with fixations, whereas the second should not accept SP in place of 
fixations.

Overall, eye movement detection lacks precise definitions and is very fragmented: 
Researchers focus on detecting certain eye movement types in isolation \cite{AgStDo16, steil2018fixation, behrens2010improved}, thus potentially 
missing important relations between them. The data sets that are assembled for such 
works, especially with VR or mobile eye tracking, are scarce and often specialised: 
E.g.\ \cite{santini2016bayesian} use mobile eye tracking, but without 
head motion and only with synthetic stimuli; 
\cite{steil2018fixation} annotate gaze target similarity and not the actual eye movements; 
\cite{john2017dataset} touches on the difficulty of understanding and annotating 
eye movements during head or body motion,
but does not explicitly define the labelled eye movement types.

The contributions of our work are as follows: We recorded and made publicly
available a data set of eye tracking recordings for dynamic real-world
360$^\circ$ video free-viewing as well as for one synthetic video clip, where
eye movements are inferred more easily.  Our data total ca.\ 3.5\,h of
recordings.
We developed a two-stage manual annotation procedure that labels (in line with
typical expert annotations) fixations, saccades, and pursuits, as well
as higher-level concepts that describe eye-head coordination (vestibular-ocular
reflex -- VOR and pursuing with head movement only) or interaction of
several ``basic'' eye movement types, such as (optokinetic) nystagmus (OKN).
We implemented this procedure by extending the open-source eye tracking data
labelling interface of \cite{agtzidis2016pursuit} for eye movement annotation
with 360$^\circ$ content.
With its help, we manually annotated a part of the collected data (ca.\ 16\%,
two representative observers per clip), which already allows for the evaluation
of algorithmic labelling approaches.
We attempted to give operational definitions to all the labelled eye movements
and provide both a theoretical and a data-driven basis for future research.
Based on the principles underlying our manual annotation, we also devised a
simple unified framework for algorithmically detecting all the eye movement
classes we defined. To the best of our knowledge, this is the first combination
of a data set and a framework to attempt systematically labelling all major
occurring eye movement types in a unified fashion.  
Our algorithm is also the first eye movement detection method that combines information
from both eye-in-head and eye-in-world frames of reference. 

\section{Related Work}

As the contributions of this work are tightly related to eye tracking set-ups
with unrestricted head rotation, we mostly focus on the works in the same
domain, including mobile and VR eye tracking.

\subsection{Data Sets and Eye Movement Annotation}

For egocentric or 360$^\circ$ content, only few data sets are available that
provide raw eye tracking data so far. Even fewer studies supply manual
annotations or develop an algorithmic detection strategy for the eye movements
in this context.  Saliency in 360$^\circ$ \cite{gutierrez2018toolbox,
cheng2018cube, nguyen2018attention} as well as egocentric
\cite{li2018sparse, lee2012discovering, polatsek2016novelty} content is gaining
popularity, and this inevitably requires the collection of eye tracking data
for 360$^\circ$ images and videos or in the mobile eye tracking scenario.
However, the data sets that are typically published provide scanpaths in the
form of sequences of ``fixations'' \cite{rai2017dataset, david2018dataset,
bolshakov2017saliency, sitzmann2018saliency}, which limits their usefulness
for eye movement research.  Also, while the frequency of the eye tracker is
not that important for saliency analyses, higher-frequency data that is
available with modern eye trackers enables much finer-grained eye movement
detection. \cite{damen2014you, santini2016bayesian}, for example, provide
the eye tracking data at 30\,Hz only.

\cite{santini2016bayesian} use mobile eye tracking, but restrict the participants'
movements with a chin bar and do not project the gaze coordinates onto the 
scene camera feed. The diversity of this data set is limited by the synthetic nature
of the stimuli.
\cite{steil2018fixation} annotate the data for their own definition of fixations, 
which does not necessarily correspond to the eye movements themselves: The annotators were
labelling sequences of dwells on real-world objects, regardless of whether these objects
were moving or whether the observer's head was in motion. 
\cite{fischer2018rtgene} use mobile eye tracking to validate a remote imaging-based eye gaze estimation
approach only, without analysing the eye movements. 
\cite{lo2017video} only capture the head rotation data without any eye tracking, assuming 
that the object at the centre of the participant's field of view is the one being looked at.
\cite{polatsek2016novelty} seem to use the term ``fixation'' interchangeably with ``gaze point''.

\cite{lowe2015visualization} designed a visualisation interface for
multi-viewer gaze similarity for 360$^\circ$ content analysis. 
This is, however, not a tool for eye movement annotation or analysis. 
\citep{john2017dataset, kothari2017gaze} manually annotated eye movements in
recordings with a wearable eye tracker during locomotion and ball catching. The
head motion was reconstructed with the data from a six-axis inertial
measurement unit.  During the annotation, the head and the eye-in-head speeds
were displayed alongside the feed from the eye and scene cameras (with the gaze
projection marked with a cross). The authors labelled fixations,
pursuits, saccades, and blinks, but no explicit eye movement definitions that
were used for manual annotation are given.  In general, this labelling method
takes into account both the eye-in-head and the eye-in-world movements, but
only implicitly -- through comparing the eye and head speeds or by inferring
the gaze point motion on the scene camera frames. This makes it harder for the
experts to understand the precise nature of the eye movement, especially when
the participant, their gaze, and the scene objects are all moving at the same
time.

In comparison to previously published works, our data set is relatively
high-frequency (120\,Hz), provides the raw (re-calibrated) eye tracking
recordings, and is (partially) manually annotated with explicit definitions of
the labelled eye movements. In addition to the typically considered set of
fixations, saccades, and pursuits, we annotate instances of VOR and OKN, as
well as pursuits of objects performed with the head only (no eye-in-head
movement).  Our annotation process takes advantage of a two-stage
pipeline, where only eye-in-head motion is considered during the first stage,
and the labelling is refined with the reference to eye-in-world motion during
the second stage. This allows the annotator to access all available information
about the recorded signal sequentially instead of all at once, thus simplifying
the procedures of the individual stages. 

\subsection{Algorithmic Detection}

Most of the algorithms so far have been developed with monitor-based experiments
in mind (due to their prevalence in research to date). Therefore, they cannot 
distinguish whether the provided gaze recordings are in the coordinate system relative
to the head (i.e.\ eye-in-head gaze) or relative to the world (eye-in-world gaze).
The frame of reference of the ensuing gaze data analysis, therefore, usually depends on the 
recording type: For wearable eye trackers (mobile or integrated into an HMD), 
eye-in-head gaze is commonly~analysed, for fixed eye trackers -- eye-in-world (e.g.\ gaze on the monitor).

The built-in algorithms for two of the most popular wearable eye trackers use
only the eye-in-head frame of reference for saccade and fixation detection. The
TobiiPro software uses a speed-based I-VT filter \cite{olsen2012tobii} when the
Tobii Pro Glasses~2 eye tracker is used. The Pupil Labs headset uses a modified
\cite{pupil-fix-det} version of the gaze dispersion-based algorithm I-DT
\cite{salvucci2000identifying} and simply handles gaze direction vectors
instead of on-screen coordinates.  \cite{david2018dataset} also use a version
of I-VT in the frame of reference of the head, and \cite{sitzmann2018saliency}
use I-DT in the frame of reference of the virtual environment.

These simplified approaches will necessarily mislabel eye movement types in the
presence of head motion, which is often present in unconstrained scenarios.  In
our data, for example, 48\% of the time the head was moving with a speed of at
least $10^\circ/s$.  Since the Tobii Pro Glasses 2 are equipped with a
gyroscope, augmented by \cite{hossain2016eye} to account for head motion (not
yet used in the software shipped to the user). The approach
of~\cite{kinsman2012ego} compensates for ego-motion by using the movement
information obtained from the scene camera (this is a more widely applicable,
but less precise approach compared to using sensor data).

In their respective approaches, \cite{anantrasirichai2016fixation} and
\cite{steil2018fixation} define fixations as maintaining gaze on an object in
the world, regardless of head movement, locomotion, and object motion.  This
definition, similar to labelling virtual object dwells, mixes up dynamic and
static eye movements and does not account for the interplay of head and eye
movements, though slightly different mechanisms are at work during coordinated
head-eye actions \cite{fang2015eye, angelaki2009vestibulo}.
\cite{steil2018fixation} use a purely image-based technique, which computes the
similarity score of the scene camera frame patches around subsequent gaze
locations with a pre-trained deep network \cite{zagoruyko2015learning}. They
assign ``fixation'' labels to gaze samples that correspond to the patches that
are similar (above a certain threshold) to the patch of the previous gaze
sample.  \cite{anantrasirichai2016fixation} combine some pre-trained deep
network-based features at the gaze location with position-derived statistics in
order to detect fixations.

Optokinetic nystagmus (OKN) is a relatively easily-detectable eye movement
pattern, and the algorithm of \cite{turu2014method} first splits the input
signal into fast and slow phases, then detecting episodes of OKN when the gaze
moves in roughly opposite directions during the fast and slow phases.

Typical papers on eye movement detection focus on a certain aspect of the data,
or even a certain eye movement type \cite{turu2014method, AgStDo16,
larsson2016smooth}. It has been noted before that not labelling some eye
movement types likely leads to poorer detection of the others (false
detections that could not be attributed to any other class)
\cite{andersson2017one}. In contrast to this, we attempted to develop a
universal eye movement labelling scheme that is based on the definitions of
the eye movements, which we provide in this work, as these can differ
wildly and unexpectedly from researcher to researcher \cite{hooge2017human,
hessels2018eye}.

\section{Data Set Collection}

Gathering a data set of eye tracking recordings for 360$^\circ$ equirectangular
videos differs from the common monitor-based experiments.  The experimental
set-up, the choice criteria for the used stimuli, as well as the way of
accounting for drifts during recordings are all influenced by the stimulus
type. We explain our choices and describe the full data collection procedure
below.

\subsection{Hardware and Software}
\label{sec:setup}

For data gathering we used the FOVE\footnote{\url{https://www.getfove.com}} virtual
reality headset with an integrated 120\,Hz eye tracker. For video presentation
we used the integrated media player of
SteamVR\footnote{\url{https://store.steampowered.com/steamvr}}, which supports
360$^\circ$ content (we used equirectangular video format). A small custom C++
program was used to handle the eye tracking recordings and store them to disk.
The data we stored for each recording includes (i) $x$ and $y$ coordinates of
the gaze point on the full 360$^\circ$ video surface in equirectangular
coordinates, (ii) the same $x$ and $y$ coordinates of the head direction, as
well as its tilt. This allowed us to disentangle the eye motion from the head
motion (computing the eye-in-head motion) and to reconstruct the gaze position
in each participant's field of view.  We also stored (as metadata) the
dimensions of the headset's field of view (in degrees and in pixels).

We kept the original sound of the presented videos.  In all clips but two it
corresponded to the environment noises (the two exceptions had silence and an
overlaid soundtrack).  Sound has a bearing on eye movements during
monitor-based video viewing \cite{coutrot2012influence}, and should affect the
viewers even more in virtual environments as noises may induce head rotation
towards video regions that would otherwise never be in the field of view.

In our experimental set-up the participants were sitting on a swivel chair with
the headset and headphone cables suspended from a hook above them. This
allowed the subjects to swivel on the chair freely, without the interference of
the cords, which could have otherwise led them to avoid head rotation.  In
addition to the discomfort of feeling the attached cables, unless those are
suspended from above, their stiffness would have likely caused the displacement
of the headset relative to the observer's head during the experiment, thus
lowering the quality of eye tracking recordings.

\subsection{Stimuli}
\label{sec:stimuli}

The collection of videos we assembled includes 14 naturalistic clips we chose
from YouTube and one synthetically generated video.  All the naturalistic data
are licensed under the Creative Commons 
license\footnote{The license used by YouTube is the more permissive version of Creative Commons -- \url{https://creativecommons.org/licenses/by/3.0/legalcode} -- and allows reuse, remix, and distribution of the original work with attribution to the original creator.}. 
We give attribution to the original creators of the content by providing the
Youtube IDs of the original videos together with our data set. The selected
clips represent different categories of scene content and context, e.g.\ static
camera, walking, cycling, or driving, as well as such properties as the content
representing an indoors or an outdoors scene, the environment being crowded or
empty, urban or mostly natural. The durations of the complete videos varied
greatly, and we decided to use a maximum of one minute per stimulus.  For each
of these clips, we extracted a continuous part of the original recording that
contained no scene cuts to preserve the immersion. The details for each video
(name, categories somewhat describing the scene, duration) are listed in
Table~\ref{tab:videos}.

\begin{table}[t!]
	\caption{Video Stimuli}
	\label{tab:videos}
	\begin{tabular}{lcccc}
		\toprule
		Video Name & Categories & Duration \\ \midrule
		01\_park & static camera, nature, empty & 1:00 \\
		02\_festival & static camera, urban, busy & 1:00 \\
		03\_drone & drone flight, urban, very high & 1:00 \\
		04\_turtle\_rescue & static camera, nature, busy & 0:38 \\
		05\_cycling & cycling, urban, busy & 1:00 \\
		06\_forest & walking, nature, empty & 1:00 \\
		07\_football & static camera, nature, busy & 1:00 \\
		08\_courtyard & static camera, urban, busy & 1:00 \\
		09\_expo & static camera, indoors, busy & 1:00 \\
		10\_eiffel\_tower & static camera, urban, busy & 0:57 \\
		11\_chicago & walking, urban, busy & 1:00 \\
		12\_driving & car driving, urban, busy & 1:00 \\
		13\_drone\_low & drone flight, urban, empty & 1:00 \\
		14\_cats & static camera, urban, busy & 0:43 \\ 
        15\_synthetic & moving target & 1:25 \\
		\bottomrule
	\end{tabular}
\end{table}

In addition, we generated one stimulus clip synthetically for a more controlled
scenario.  The circular gaze target we used for this part of the experiment
followed the recommendations of \cite{thaler2013best} in order to improve
fixation stability. It measured two degrees of visual angle in diameter and was
displayed in white on a black background.
For simplicity, we neglected the idiosyncrasies of the equirectangular format
for the stimulus generation here, as the target always stayed close to the
equator of the video, meaning that shape distortions would be small.

The synthetic clip we generated consisted of five phases. Each phase had a short 
instruction set displayed (for ca.\ 7\,s)
before the fixation gaze target appeared.
The first four phases lasted 10\,s \mbox{after} the stimulus appeared and were designed (together
with their respective instructions) to induce (i)~eye movements that are
typically seen in controlled lab settings: fixations, saccades, and
smooth pursuit, without excessive head motion, 
(ii)~VOR with voluntary head motion while maintaining a fixation on a stationary target,
(iii)~``natural'' long pursuit, without any
additional instructions (an arbitrary combination of body or head rotation, VOR, and smooth pursuit),
where the target moved with a constant speed of 15$^\circ/s$, covering 150$^\circ$,
and (iv)~a special combination of VOR and 
smooth pursuit, when the eyes are relatively stationary inside the
head, but the gaze keeps track of a moving target. We refer to the latter type of eye-head
coordination as ``\emph{head pursuit}''. 
During the fifth phase, OKN was induced by targets rapidly moving for a short
period of time (at 50$^\circ/s$ symmetrically around the centre of the video), 
disappearing, and then repeating the motion,
covering 25$^\circ$ on each pass.
Both left-to-right and right-to-left moving targets were displayed with a brief 2.5\,s pause
between the sequences of same-direction target movement (5\,s each). 

\subsection{Experimental Procedure}

In order to be able to detect and potentially compensate for eye tracking
quality degradation, we added a stationary fixation target at the beginning
(for 2\,s) and the end (for 5\,s) of each video clip. Overall, the 15 videos
have a cumulative duration of ca.\ 17 minutes including these fixation targets.
The recording process was split into three sessions for each participant.
During the first and the second sessions, 7 naturalistic videos were presented
in succession. The last session only included the synthetic video. 
The participants could have an arbitrary-length break between the sessions.
The eye tracker was calibrated through the headset's built-in routine shortly
before every recording session.  We then empirically and informally validated
the calibration using the FOVE sample Unity
project\footnote{\url{https://github.com/FoveHMD/FoveUnitySample}} where the
participant's gaze is visualised. If the quality was deemed insufficient, the
calibration procedure was repeated. We accounted for eye tracking drifts
between recordings of the same session by performing a one-point re-calibration
with the fixation target at the beginning of each video.

The naturalistic videos were presented in a pseudo-random order (same for all
subjects); the synthetic clip was presented last not to prompt the
observers to think about the way they moved their eyes before it was necessary.
If the participant at any point was feeling unwell, the recording was
interrupted.  Afterwards, a new calibration was performed, the unfinished video
was skipped, and the recording procedure was continued from the next clip.

Overall, we recorded gaze data of 13 subjects, and the number of recordings
per stimulus video clip was between 11 and 13 (12.3 on average), which
amounts to ca.~3.5\,h of eye tracking data in total. 

\section{Manual Annotation}
\label{sec:labelling_method}

When working with 360$^\circ$ equirectangular videos, the natural visualisation
of the recording space is the camera (or the observer's head) placed at
the centre of a sphere that is covered by the video frame pixels.
Computationally, this directly matches the equirectangular video representation,
where the $x$ and $y$ coordinates on the video surface are linearly mapped 
to the spherical coordinates of this sphere (longitude and latitude, respectively).
Since the field of view is limited (up to 100$^\circ$ in our HMD),
the observers will use head rotation (as in everyday life) to explore 
their surroundings, so this aspect of the viewing behaviour
needs to be accounted for both in the definitions of the eye movements and 
the annotation procedure.

\subsection{Definitions}
\label{sec:definitions}

In order to fully describe the interplay of the movement of the head and the eyes themselves,
we cannot assign just a single eye movement label to every gaze sample,
since the underlying process may differ when eye-head coordination is involved.
Therefore, we used two labels for each gaze sample, to which we refer as \emph{primary} and
\emph{secondary} labels. 
Following the recommendations of \cite{hessels2018eye}, we defined the eye
movements that we annotated below to avoid potential confusion in terminology.
As researchers can disagree on the nature and purpose of various eye movements
\cite{hessels2018eye}, we do not argue that the ones we used for this work are
the ultimately correct ones, but we hope that this would provide a starting
point for further refinement and investigation.  We did not include
post-saccadic oscillations or microsaccades in our annotations as the wearable
eye tracker frequency and precision did not permit their confident localisation
by the annotator.

\emph{Primary} label is necessarily assigned to \emph{all} gaze samples, and can be one
of the following:

\begin{itemize}[leftmargin=*,noitemsep,topsep=0pt,parsep=0pt,partopsep=0pt, after=\vspace{3pt}]
\item \textit{Fixation}: A period of time where no movement of the eye inside the head is triggered by
retinal input (but may e.g.\ reflexively compensate for head motion).
\item \textit{Saccade}: High-speed ballistic movement of the eye to
shift the point of regard, thus bringing a new (part of an) object 
onto the fovea (including adjusting the gaze position to match the tracked object via
catch-up saccades during pursuit, or similar).
\item \textit{Smooth pursuit (SP)}: A period of time during which the eyes are in motion inside the head and a moving
(in world coordinates, relative to the observer) target is being foveated. 
\item \textit{Noise}: Even though noise is not an actual eye movement type, we accumulate 
blinks, drifts, tracking loss, and physiologically impossible eye ``movements'' under this one name.
\end{itemize}

The \emph{secondary} labels describe in more detail how the primary eye movements were executed and
are mostly a consequence of head motion (except for OKN). The following labels are possible:
\begin{itemize}[leftmargin=*,noitemsep,topsep=0pt,parsep=0pt,partopsep=0pt, after=\vspace{3pt}]

\item \textit{Vestibulo-ocular reflex (VOR)}: A period of time when the eyes are 
compensating for head motion and stabilising the foveated area.
\item \textit{Optokinetic nystagmus (OKN) or nystagmus}: Sawtooth-like eye movement patterns, which
are composed of fast saccadic parts alternating with slow stabilisation parts. We assigned the label
of OKN to all such patterns, though it has to be noted that some of the OKN labels correspond to 
nystagmus, e.g.\ when a participant is observing a blank part of the synthetic stimulus while
simultaneously turning the head, so the reflexive movement is not 
actually triggered by the visual input.
\item \textit{VOR + OKN}: This is a combination of the two previous categories:
The eye signal exhibits a sawtooth pattern during head rotation.
\item \textit{Head pursuit}: A period of time where a pursuit of a moving target is performed only
via head motion, with the gaze direction within the head relatively constant.
\end{itemize}

Unlike the \emph{primary}, the \emph{secondary} label can easily be unassigned even in large windows of gaze samples 
(e.g.\ foveating a stationary or moving object in the scene without head motion).

\subsection{Labelling Procedure}

To thoroughly describe the labelling process, we focus primarily on the information
that was available to the manual annotator. We implemented 
a two-stage annotation pipeline, with stages corresponding to different frames of reference 
(for the visualised gaze speed and coordinates), sets of assigned labels,
and projections used for the scene content display.
We refer to these stages (or modes of operation) as \textit{field of view} and \textit{eye+head}.

In the \textit{field of view (FOV)} mode, the annotator is presented with the view of the 
scene that is defined by the corresponding head rotation of the subject (the size of the 
visualised video patch roughly corresponds to the field of view that the participant had in the 
VR headset). This view corresponds to the frame of reference that moves together
with the participant's head and allows us to see the actual visual stimulus that was
perceived by the participant.

In the \textit{eye+head (E+H)} mode, the full equirectangular video frame is
presented to the annotator. Visualising gaze locations in this view enables the
annotator to see the combination of the head and eye movement, which corresponds
to the overall gaze in the world (or 360$^\circ$ camera, to be more precise) frame of reference.

\begin{figure}[t!]
	\centering
	\begin{subfigure}{\linewidth}
		\includegraphics[width=\linewidth]{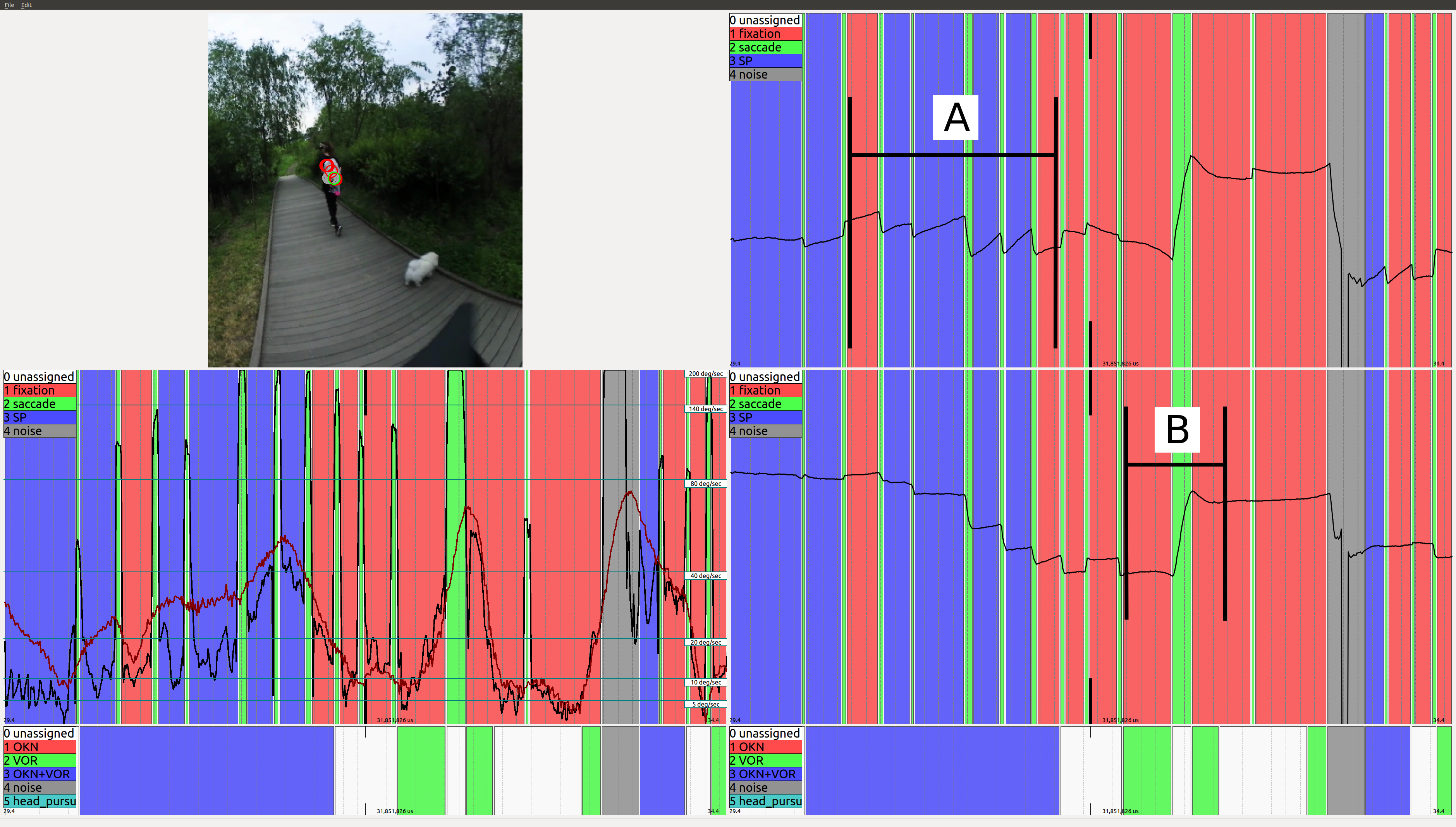}
		\caption{FOV mode}
		\label{fig:screenshot_fov}
	\end{subfigure}
		
	\begin{subfigure}{\linewidth}
		\includegraphics[width=\linewidth]{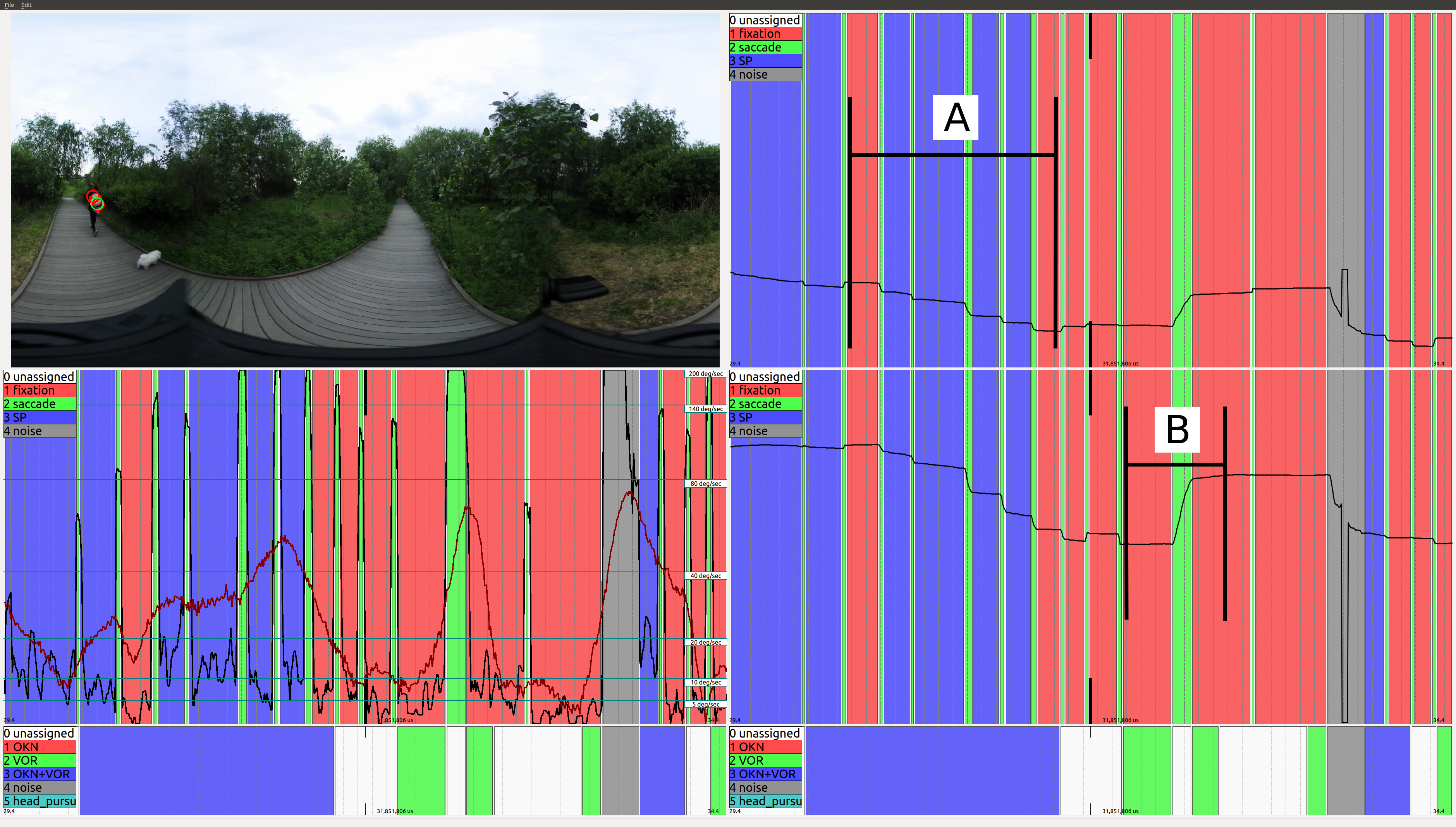}
		\caption{E+H mode}
		\label{fig:screenshot_eye_head}
	\end{subfigure}
	\caption{
		Schematic  of field-of-view (a) and eye+head (b) operation modes.
Differences in the~patterns of gaze coordinates and speeds allow for improved annotation.
``A'' and ``B'' marks are for reference 
     only (not shown during annotation). Coloured intervals correspond
     to different primary (on three large panels) and secondary 
     (bottom) labels.}
	\label{fig:screenshot}
\end{figure}

In both operation modes, the currently considered gaze sample as well as previous and future gaze
locations (up to 100\,ms) are overlaid onto the displayed video surface.
In addition, the plots of the $x$ and $y$ gaze coordinates over time,
as well as the plot of both the eye and the head speeds are presented 
(see Figure~\ref{fig:screenshot_fov} and \ref{fig:screenshot_eye_head} for the FOV and E+H mode examples). 
The coordinate systems used for these plots, however, differ 
between the two modes: In the FOV mode, the gaze coordinates and the speed of gaze
are reported in the \emph{head}-centred coordinate system, whereas in the E+H mode,
the coordinates and the speed in the \emph{world} coordinate system are visualised.
This way, the FOV representation provides the annotator with the eye motion information within
the eye socket, while the E+H representation is responsible for highlighting the absolute 
movement of the foveated objects, which is necessary for determining the precise label type,
e.g.\ distinguishing between fixations and pursuits.

The manual annotator began (i.e.\ \emph{the first stage}) 
with the FOV operation mode and assigned all
primary eye movement labels without taking head motion into account:
Ballistic eye-in-head motion would correspond to saccades,
relatively stationary (in the coordinate system of the head) gaze direction 
-- to fixations, smoothly shifting gaze position -- to pursuits 
(provided that a correspondingly moving target exists in the scene), etc.
To speed up the process, we pre-labelled saccades with 
the I-VT algorithm of \cite{salvucci2000identifying},
applied in the FOV 
coordinates (instead of the coordinates of the full equirectangular video)
with a speed threshold of $140^\circ/s$. 
The labeller then went through each recording, correcting saccade limits or inserting
missed ones, assigning fixation, SP, and noise labels, inserting new events where necessary. 
OKN was labelled in this stage as well, because the sawtooth pattern of 
eye coordinates was more visible
without the head motion effects.

After the annotator felt confident about the first labelling stage results, 
\emph{the second stage} would begin: The annotator went 
through the video again, this time -- in the E+H operation mode. 
On the second pass, the previously assigned primary labels were visible 
and needed to be re-examined in the context of the eye-head coordination,
with respective additions of the secondary labels:
\begin{itemize}[leftmargin=*,noitemsep,topsep=0pt,parsep=0pt,partopsep=0pt]

\item \textit{SP to fixation}: If the primary SP label of the first stage
corresponded to the foveation of a stationary (in world coordinates) target, the
label was changed to a fixation, and a matching VOR episode was added to the secondary
labels. If the SP episode in question
belonged to an OKN episode, the respective part of the latter was re-assigned to
the OKN+VOR class.
\item \textit{Fixation + head pursuit}: If the primary fixation label of the first stage (i.e. little to no
        movement of the eye within its socket) corresponded to 
following a moving (in world coordinates) target, the secondary ``head
pursuit'' label was added. 
\item If the primary SP label was maintained in the second stage in the presence of
head motion, a VOR episode was added to the secondary labels.
\end{itemize}

The annotation was performed by an experienced eye movement researcher
(one of the authors), who first annotated five minutes of pilot data  
in order to familiarise himself with the procedure and the interface;
ambiguities were discussed with the co-authors.
Labelling a single recording (of about a minute of gaze data) 
took between 45\,min and 1\,h.  In total, our annotations cover about 16\% of the
data (two recordings per stimulus clip) and amount to ca.\ 33\,min. 

\subsection{Labelling Tool}
\label{sec:gui}

Some previous works prefer to hide the stimulus from the 
annotator \cite{larsson2013detection} not to bias the rater's expectation of which
eye movements are more likely or possible with a given stimulus. 
We argue, however, that since we are more interested in accurate labelling than
in stimulus-agnostic distinguishability of the eye movements, providing all the 
available relevant information is an essential step. Without the video frames, 
it would be impossible to distinguish e.g.\ pursuits and drifts.

In order to implement our manual annotation pipeline we significantly extended 
the publicly available\footnote{\url{https://www.michaeldorr.de/gta-vi/}} 
hand-labelling tool of \cite{agtzidis2016pursuit},
adding the support for simultaneous primary and secondary label assignment and 
the field-of-view (FOV) operation mode, where the displayed video 
\mbox{``re-enacts''} the participant's head movements 
during the recording session
(see Figure~\ref{fig:screenshot_fov}).

For both stages of the labelling process, our interface included six
panels (see interface examples in Figure~\ref{fig:screenshot}):
The top left panel displays the video (either the FOV representation or the full 
equirectangular frame) overlaid with gaze samples.
The panel below it displays the speed of gaze (in black)
in the respective coordinate system -- head-centric for the FOV mode and the video 
coordinate system in the E+H mode -- and the speed of the head movement (in red). 
The two top panels on the right visualise the $x$ and the $y$ gaze coordinates over time
(again, the coordinate system depends on the operation mode). 
The speed and coordinate panels colour-code the time intervals according 
to the assigned primary label.
The two bottom panels are identical and serve the purpose of visualising
the secondary labels. This information is duplicated in order to give the annotator 
the possibility to easily adjust the VOR and head pursuit intervals based on both 
the head speed plot and the plot of the gaze coordinates (e.g.\ verifying that 
the gaze direction is relatively constant in the world coordinates, but the head is turning).

Despite the multitude of panels in the interface, only a subset was used to make the
vast majority of decisions: Gaze coordinate panels were mostly sufficient for primary 
and secondary label assignment. The speed and video panels were referred to in case
of uncertainty.

Figure~\ref{fig:screenshot} also illustrates the differences of gaze patterns in the two 
representations we use. For instance, the sawtooth pattern that can be observed in FOV view 
close to the beginning of the displayed gaze sequence changes shape in the E+H mode and becomes
rather step-shaped (regions marked with ``A'' in the figure). 
Also note how the head and eye movements cancel out each other during a fixation
that is combined with VOR (regions marked with ``B'', corresponding to the VOR labels on the two 
bottom panels in Figure~\ref{fig:screenshot}): The speeds are almost equal 
in the FOV mode (Figure~\ref{fig:screenshot_fov}), and the eye in the world coordinate system 
is almost stationary (Figure~\ref{fig:screenshot_eye_head}).

\section{Eye Movement Detection Algorithm}

We now describe a rule-based eye movement classification algorithm that 
is almost a direct formalisation of the eye movement definitions we consider in 
Section~\ref{sec:definitions}. 
It assigns primary and secondary labels to every gaze sample (potentially ``unassigned'' for the secondary labels) 
by analysing the same gaze and head movement information that was 
available to the manual annotator.

We first detected the saccades by analysing the E+H speeds with the two-threshold algorithm
of \cite{dorr2010variability}, which avoids false detections while maintaining 
high recall by requiring each saccade to 
have a peak gaze speed of at least 150$^\circ/s$, but all surrounding samples with 
speeds above $35^\circ/s$ are also added to the detected episode.
We did not use the FOV speed of gaze as it is influenced by head motion and can easily
reach speeds above $100^\circ$/s when the eyes compensate for fast large-amplitude
head rotations. 

Afterwards, blinks were detected by finding the periods of lost tracking and extending them
to include saccades that were detected just prior to or just after these periods, as long as
the saccades were not farther than 40\,ms from the samples with lost tracking.

We then split the remaining intersaccadic intervals into non-overlapping
windows of 100\,ms and classified each such interval independently.
For this, we calculated the speeds of the head and the eye (relative to the head
and the world) as the distance covered from the beginning 
to the end of the window divided by its duration.

To formalise the concepts of ``stationary'' and ``moving'' head cases, we used a speed threshold of $7^\circ/s$.
For the gaze speeds, we applied the low and the high thresholds of $10^\circ/s$ and
$65^\circ/s$, respectively 
(both for the eye-in-head and the eye-in-world speeds) in order to distinguish slow,
medium, and fast movements.
As gaze stability decreases with head motion \cite{ferman1987human}, we scaled the gaze speed
thresholds according to the speed of the head: 
$thd_\text{scaled} = (1 + {v_\text{head}}/{60}) * thd$, 
where $60^\circ/s$ is the ``reference'' speed of the head. 
This means that if the head was moving at e.g.\ 30$^\circ/s$, the gaze speed
thresholds were increased by 50\%.

A fixation was always labelled when the E+H speed was below the low gaze speed
threshold. If the head speed was above the corresponding low threshold, 
a secondary VOR label was assigned.

Pursuit-type eye movement labels were assigned when the E+H speed was between the low and
the high gaze speed thresholds, unless the eye-in-head speed was above the high threshold
(in which case, a noise label was assigned).
However, there are different label combinations possible
here: (i) \emph{Head pursuit} in combination with the primary label of \emph{fixation} was
assigned when the FOV (eye-in-head) speed was below the low threshold and the head speed 
was above its own low threshold;
otherwise, (ii) \emph{smooth pursuit} in combination with \emph{VOR} was detected when the head speed was
above the low threshold, which implied that the head and the eyes were working in tandem 
(presumably, to follow a moving object);
(iii) \emph{smooth pursuit} without any secondary eye movement type was assigned
when the head speed was below its low threshold, meaning that the eyes did not have to compensate 
for the head~movement.

\begin{table}[t!]
	\caption{Threshold Values}
	\label{tab:parameters}
	\begin{tabular}{lccc}
		\toprule
		Name & Used for & Threshold & Optimised \\ \midrule
		$\theta_{sacc}^{low}$ & saccades & 35$^\circ$/s & \checkmark \\ [1ex]
		$\theta_{sacc}^{high}$ & saccades & 150$^\circ$/s & \checkmark \\ [1ex]
		$\theta_{gaze}^{low}$ & fix., SP, VOR, head purs. & 10$^\circ$/s & \checkmark \\ [1ex]
		$\theta_{gaze}^{high}$ & fix., SP, VOR, head purs. & 65$^\circ$/s & \checkmark \\ [1ex]
		$\theta_{head}^{low}$ & VOR, head purs. & 7$^\circ$/s & - \\ [1ex]
		$\theta_{head}^{high}$ & scaling $\theta_{gaze}^{\{low, high\}}$ & 60$^\circ$/s & - \\ [1ex]
		\bottomrule
	\end{tabular}
\end{table}

For the samples that did not fall into any of the previously listed categories it was then known
that they had very high speed but~were assumed not to be a part of any saccade 
(since saccades were detected already). Consequently, the noise label was assigned.

Overall, our approach uses five speed thresholds (plus a scaling parameter), and thus we
refer to our algorithm as I-S$^5$T, \emph{identification by five speed thresholds}.
An overview of the parameters is given in Table~\ref{tab:parameters}:
two thresholds for saccade detection, two to quantise eye speeds (scaled by
head speed), and one to determine if the head is moving sufficiently to justify
a potential VOR label. The used values for the first four of these were
optimised using a grid-search procedure on the entire annotated data set,
as we were interested in determining the upper-bound of what can be achieved with
a simple detection algorithm in this relatively complex task, 
rather than in finding a well-transferable set of
precise threshold values.

We also implemented an algorithm for detecting OKN (or nystagmus),
with its sawtooth pattern of gaze coordinates. This pattern is easier to detect 
in the FOV gaze data as it often occurred during high-amplitude head motion in our data. The idea
behind our detector is similar to \cite{turu2014method}, but uses the already detected saccades for
segmenting the recordings into slow and fast phases,
instead of finding the maxima and minima in the speed signal.
An OKN is detected when the overall direction of gaze movement 
during an intersaccadic interval is roughly opposite (angle $\ge90^\circ$) 
to the direction of the adjacent saccades, whereas the two
neighbouring saccades are roughly collinear (angle $\le70^\circ$). 
In case of an already assigned VOR label, OKN+VOR is labelled
instead.

\section{Results and Discussion}

We publicly provide the entire collected eye tracking data set and its (partial) manual
annotation, together with the video clips that were used as stimuli and 
the implementation of the annotation tool and the I-S$^5$T algorithm on the 
project page: \url{https://web.gin.g-node.org/ioannis.agtzidis/360_em_dataset}. 

The collected hand-labelled data make it possible to examine in detail the eye
movement patterns and typical behaviours that observers exhibit when viewing
dynamic 360$^\circ$ content.  The assigned primary eye movement labels consist
of 74.9\% fixations (4193 events), 10.5\% saccades (3964 events), 9.9\% SP (552
events), and 4.7\% noise (553 events). The secondary eye movement
labels include 28.0\% VOR (1825 events), 15.9\% of a combination of OKN+VOR
(295 events), 0.8\% OKN without VOR (21 events), and 1.4\% head pursuit (52
events).  We believe that this is the first data set that addresses the
eye movement strategies in 360$^\circ$ video viewing for such a large spectrum
of eye movement classes at the same time.  Our data can serve as basis for
further gaze behaviour analysis and gaze event detection algorithm testing.

\subsection{Automatic Classification Quality}
\label{sec:evaluation}

\begin{table}[t!]
\caption{Classification Performance on the Test Set}
\label{tab:results}
\begin{tabular}{clcccccc}
\toprule
&  & \multicolumn{3}{c}{Sample F1} & \multicolumn{3}{c}{Event F1} \\ \midrule
 & EM type & Comb. & FOV & E+H & Comb. & FOV & E+H\\ \midrule
\parbox[t]{0.5mm}{\multirow{4}{*}{\rotatebox[origin=c]{90}{\textit{Primary}}}} & Fixation & 0.911 & 0.867 & 0.900 & 0.897 & 0.808 & 0.890 \\ 
& Saccade  & 0.813 & 0.737 & 0.813 & 0.899 & 0.865 & 0.899 \\ 
& SP       & 0.381 & 0.128 & 0.362 & 0.288 & 0.153 & 0.293 \\ 
& Noise    & 0.758 & 0.743 & 0.758 & 0.744 & 0.729 & 0.742 \\ \midrule 
 \parbox[t]{0.5mm}{\multirow{4}{*}{\rotatebox[origin=c]{90}{\textit{Secondary}}}} & OKN      & 0.205 & -     & -     & 0.085 & -     & -     \\ 
& VOR      & 0.600 & -     & -     & 0.636 & -     & -     \\ 
& OKN+VOR  & 0.664 & 0.614 & 0.647 & 0.577 & 0.626 & 0.620 \\ 
& Head Purs.& 0.546 & -   & -     & 0.204 & -     & -     \\
\bottomrule
\end{tabular}
\end{table}

To evaluate the performance of our algorithmic event detection as well as to 
explain the benefits of utilising the data from both the eye and the head 
tracking, we compared the performance of our algorithmic detector I-S$^5$T against two
versions of the same algorithm: one that only uses the speed of the eye within
the head (e.g.\ directly applicable to mobile eye tracking data), the other -- E+H gaze data (e.g.\ in
HMD recordings, if additional data were discarded) instead of a combination of
all available movement readouts. We split our ground truth data into a training and a test set,
each containing one manually annotated eye tracking recording for each video. 
The sets of recorded participants in training and test sets do not intersect. 
For all algorithm versions, we selected the gaze speed 
thresholds (i.e.\ head speed threshold was not optimised) with a similar grid-search 
optimisation procedure on the training set -- first, the two thresholds for saccade detection were 
jointly optimised, then the remaining two gaze speed thresholds.

We refer to the algorithm versions as (i) \emph{combined} for the ``main'' proposed version -- the I-S$^5$T algorithm --
that used both the eye-in-head and eye-in-world speeds, as well as head speed for 
threshold scaling, (ii) \emph{FOV} for the 
version that used the eye-in-head gaze speed only, and  (iii) \emph{E+H} for the one
that only used the eye-in-world speeds. Of course, the FOV and E+H versions did not 
detect the combinations of head and eye movements, so the secondary labels of
VOR and head pursuit are not assigned. OKN detection is possible, however.
Since there was much more OKN+VOR than pure OKN in our data, whenever OKN was 
detected based on the FOV or E+H algorithm versions, an OKN+VOR label was 
assigned.

We evaluated all three algorithm versions on the manually labelled 
test set. Table~\ref{tab:results} contains the sample- and event-level 
evaluation measures (in the form of F1 scores) for our approaches. 
Event-level evaluation follows the procedure of \cite{hooge2017human}.

All three algorithms achieve relatively high F1 scores for fixation and saccade detection,
with the FOV version yielding substantially lower scores. 
This indicates that saccades can be easily confused with the eyes compensating for 
the head movement.
The difference is even more pronounced for SP detection, with the FOV version 
of the algorithm lagging far behind. The differences
between the E+H version and the ``combined'' versions are generally very
small for the primary eye movement classes (fixations, saccades, SP, and noise), 
with the combined variant achieving marginally higher scores. 
For the secondary labels, only the version that combined eye-in-head and 
eye-in-world speeds was able to detect the full spectrum of the defined eye movements,
as most of the secondary labels require the knowledge of both the eye and the head movement
information. OKN detection was comparable across the board.

Our evaluation has demonstrated that eye movement classification algorithms 
could benefit from using all the available information about head and gaze 
in every frame of reference. This is especially important for distinguishing 
eye movements driven by the retinal input (e.g.\ smooth pursuit) and other sensory
intakes (e.g.\ VOR), 
which is supported by the definitions of the eye movements that we introduced in
Section~\ref{sec:definitions}. Those necessarily entail that using either eye-in-head or eye-in-world 
coordinate systems exclusively does not allow distinguishing even all the primary 
eye movements from one another: E.g.\ to differentiate between fixation + VOR and SP, the eye-in-world speeds are required; to discriminate between 
fixation + head pursuit and SP labels, however, the eye-in-head coordinates
are critical. 
These observations are particularly relevant for wearable eye tracker scenarios,
as gaze coordinates are often reported in the FOV only, which corresponds to the worst-performing version
of our algorithm (despite parameter optimisation).
In this set-up, additional classification power can be gained by incorporating 
head motion information, e.g.\ from a gyroscope \cite{hossain2016eye} or from the field camera images \cite{kinsman2012ego}.

In general, using fixed thresholds (despite their scaling with gaze speed, as in I-S$^5$T)
is not as flexible as the adaptive thresholds human annotators implicitly use, which
depend on the noise level, for example. Experts also take into account a much larger
context of gaze movement for each decision (compared to 100\,ms windows in our approach).
Expanding the analysis context for the algorithms also results in improved performance \cite{startsev2018cnn}. 
Additionally, the eye movements' correspondence to 
the motion of the video objects  
is ignored by our algorithm,
but is essential for accurately detecting tracking eye movements (and readily available 
to human annotators). 
The labels of our algorithm could be further refined 
using object tracking techniques or performing gaze target similarity analysis
as in \cite{steil2018fixation}.

\section{Conclusions}
In this paper, we aimed to provide a starting point for comprehensive eye movement
classification in an unrestrained head setting. 
To this end, we selected a very generic stimulus domain (naturalistic 360$^\circ$ video),
where we can, however, retain auxiliary information such as precise head rotation.
We collected a data set of eye tracking recordings for thirteen observers 
and manually annotated a representative part of it. We also presented a simple 
rule-based eye movement classification algorithm, which we optimised and tested in 
different settings, arguing that utilising both eye-in-head and eye-in-world 
statistics is necessary for the correct identification of eye movement classes.
To the best of our knowledge, this is the first attempt to fully label the eye
movement types with freely moving head in an immersive 360$^\circ$ paradigm.
This data set and algorithm may serve as a basis to further improve both the 
theoretical and the practical foundations of eye movement detection in the real world.

\begin{acks}
This research was supported by the Elite Network Bavaria, funded
by the Bavarian State Ministry of Science and the Arts.
\end{acks}

\bibliographystyle{ACM-Reference-Format}
\bibliography{paper}


\begin{thebibliography}{44}


\ifx \showCODEN    \undefined \def \showCODEN     #1{\unskip}     \fi
\ifx \showDOI      \undefined \def \showDOI       #1{#1}\fi
\ifx \showISBNx    \undefined \def \showISBNx     #1{\unskip}     \fi
\ifx \showISBNxiii \undefined \def \showISBNxiii  #1{\unskip}     \fi
\ifx \showISSN     \undefined \def \showISSN      #1{\unskip}     \fi
\ifx \showLCCN     \undefined \def \showLCCN      #1{\unskip}     \fi
\ifx \shownote     \undefined \def \shownote      #1{#1}          \fi
\ifx \showarticletitle \undefined \def \showarticletitle #1{#1}   \fi
\ifx \showURL      \undefined \def \showURL       {\relax}        \fi
\providecommand\bibfield[2]{#2}
\providecommand\bibinfo[2]{#2}
\providecommand\natexlab[1]{#1}
\providecommand\showeprint[2][]{arXiv:#2}

\bibitem[\protect\citeauthoryear{Agtzidis, Startsev, and Dorr}{Agtzidis
  et~al\mbox{.}}{2016a}]%
        {agtzidis2016pursuit}
\bibfield{author}{\bibinfo{person}{Ioannis Agtzidis}, \bibinfo{person}{Mikhail
  Startsev}, {and} \bibinfo{person}{Michael Dorr}.}
  \bibinfo{year}{2016}\natexlab{a}.
\newblock \showarticletitle{In the pursuit of (ground) truth: A hand-labelling
  tool for eye movements recorded during dynamic scene viewing}. In
  \bibinfo{booktitle}{{\em 2016 IEEE Second Workshop on Eye Tracking and
  Visualization (ETVIS)}}. \bibinfo{pages}{65--68}.
\newblock
\showDOI{%
\url{https://doi.org/10.1109/ETVIS.2016.7851169}}


\bibitem[\protect\citeauthoryear{Agtzidis, Startsev, and Dorr}{Agtzidis
  et~al\mbox{.}}{2016b}]%
        {AgStDo16}
\bibfield{author}{\bibinfo{person}{Ioannis Agtzidis}, \bibinfo{person}{Mikhail
  Startsev}, {and} \bibinfo{person}{Michael Dorr}.}
  \bibinfo{year}{2016}\natexlab{b}.
\newblock \showarticletitle{Smooth pursuit detection based on multiple
  observers}. In \bibinfo{booktitle}{{\em Proceedings of the Ninth Biennial ACM
  Symposium on Eye Tracking Research \& Applications}} {\em
  (\bibinfo{series}{ETRA '16})}. \bibinfo{publisher}{ACM},
  \bibinfo{address}{New York, NY, USA}, \bibinfo{pages}{303--306}.
\newblock


\bibitem[\protect\citeauthoryear{Anantrasirichai, Gilchrist, and
  Bull}{Anantrasirichai et~al\mbox{.}}{2016}]%
        {anantrasirichai2016fixation}
\bibfield{author}{\bibinfo{person}{N. Anantrasirichai}, \bibinfo{person}{I.~D.
  Gilchrist}, {and} \bibinfo{person}{D.~R. Bull}.}
  \bibinfo{year}{2016}\natexlab{}.
\newblock \showarticletitle{Fixation identification for low-sample-rate mobile
  eye trackers}. In \bibinfo{booktitle}{{\em 2016 IEEE International Conference
  on Image Processing (ICIP)}}. \bibinfo{pages}{3126--3130}.
\newblock
\showISSN{2381-8549}
\showDOI{%
\url{https://doi.org/10.1109/ICIP.2016.7532935}}


\bibitem[\protect\citeauthoryear{Andersson, Larsson, Holmqvist, Stridh, and
  Nystr{\"o}m}{Andersson et~al\mbox{.}}{2017}]%
        {andersson2017one}
\bibfield{author}{\bibinfo{person}{Richard Andersson}, \bibinfo{person}{Linnea
  Larsson}, \bibinfo{person}{Kenneth Holmqvist}, \bibinfo{person}{Martin
  Stridh}, {and} \bibinfo{person}{Marcus Nystr{\"o}m}.}
  \bibinfo{year}{2017}\natexlab{}.
\newblock \showarticletitle{One algorithm to rule them all? An evaluation and
  discussion of ten eye movement event-detection algorithms}.
\newblock \bibinfo{journal}{{\em Behavior Research Methods\/}}
  \bibinfo{volume}{49}, \bibinfo{number}{2} (\bibinfo{date}{01 Apr}
  \bibinfo{year}{2017}), \bibinfo{pages}{616--637}.
\newblock
\showISSN{1554-3528}
\showDOI{%
\url{https://doi.org/10.3758/s13428-016-0738-9}}


\bibitem[\protect\citeauthoryear{Angelaki}{Angelaki}{2009}]%
        {angelaki2009vestibulo}
\bibfield{author}{\bibinfo{person}{D.E. Angelaki}.}
  \bibinfo{year}{2009}\natexlab{}.
\newblock \showarticletitle{Vestibulo-Ocular Reflex}.
\newblock In \bibinfo{booktitle}{{\em Encyclopedia of Neuroscience}},
  \bibfield{editor}{\bibinfo{person}{Larry~R. Squire}} (Ed.).
  \bibinfo{publisher}{Academic Press}, \bibinfo{address}{Oxford},
  \bibinfo{pages}{139 -- 146}.
\newblock
\showISBNx{978-0-08-045046-9}
\showDOI{%
\url{https://doi.org/10.1016/B978-008045046-9.01107-4}}


\bibitem[\protect\citeauthoryear{Barz}{Barz}{2015}]%
        {pupil-fix-det}
\bibfield{author}{\bibinfo{person}{Michael Barz}.}
  \bibinfo{year}{2015}\natexlab{}.
\newblock \bibinfo{title}{PUPIL fixation detection}.
\newblock
  \bibinfo{howpublished}{\url{https://github.com/pupil-labs/pupil/blob/master/pupil_src/shared_modules/fixation_detector.py}}.
    (\bibinfo{year}{2015}).
\newblock


\bibitem[\protect\citeauthoryear{Behrens, MacKeben, and
  Schr{\"o}der-Preikschat}{Behrens et~al\mbox{.}}{2010}]%
        {behrens2010improved}
\bibfield{author}{\bibinfo{person}{Frank Behrens}, \bibinfo{person}{Manfred
  MacKeben}, {and} \bibinfo{person}{Wolfgang Schr{\"o}der-Preikschat}.}
  \bibinfo{year}{2010}\natexlab{}.
\newblock \showarticletitle{An improved algorithm for automatic detection of
  saccades in eye movement data and for calculating saccade parameters}.
\newblock \bibinfo{journal}{{\em Behavior Research Methods\/}}
  \bibinfo{volume}{42}, \bibinfo{number}{3} (\bibinfo{date}{01 Aug}
  \bibinfo{year}{2010}), \bibinfo{pages}{701--708}.
\newblock
\showISSN{1554-3528}
\showDOI{%
\url{https://doi.org/10.3758/BRM.42.3.701}}


\bibitem[\protect\citeauthoryear{Bolshakov, Gracheva, and Sidorchuk}{Bolshakov
  et~al\mbox{.}}{2017}]%
        {bolshakov2017saliency}
\bibfield{author}{\bibinfo{person}{Andrey Bolshakov}, \bibinfo{person}{Maria
  Gracheva}, {and} \bibinfo{person}{Dmitry Sidorchuk}.}
  \bibinfo{year}{2017}\natexlab{}.
\newblock \showarticletitle{How many observers do you need to create a reliable
  saliency map in VR attention study?}. In \bibinfo{booktitle}{{\em Abstract
  Book of the European Conference on Visual Perception (ECVP)}}.
\newblock


\bibitem[\protect\citeauthoryear{Cheng, Chao, Dong, Wen, Liu, and Sun}{Cheng
  et~al\mbox{.}}{2018}]%
        {cheng2018cube}
\bibfield{author}{\bibinfo{person}{Hsien-Tzu Cheng}, \bibinfo{person}{Chun-Hung
  Chao}, \bibinfo{person}{Jin-Dong Dong}, \bibinfo{person}{Hao-Kai Wen},
  \bibinfo{person}{Tyng-Luh Liu}, {and} \bibinfo{person}{Min Sun}.}
  \bibinfo{year}{2018}\natexlab{}.
\newblock \showarticletitle{Cube Padding for Weakly-Supervised Saliency
  Prediction in {\ang{360}} Videos}. In \bibinfo{booktitle}{{\em The IEEE
  Conference on Computer Vision and Pattern Recognition (CVPR)}}.
\newblock


\bibitem[\protect\citeauthoryear{Coutrot, Guyader, Ionescu, and
  Caplier}{Coutrot et~al\mbox{.}}{2012}]%
        {coutrot2012influence}
\bibfield{author}{\bibinfo{person}{Antoine Coutrot}, \bibinfo{person}{Nathalie
  Guyader}, \bibinfo{person}{Gelu Ionescu}, {and} \bibinfo{person}{Alice
  Caplier}.} \bibinfo{year}{2012}\natexlab{}.
\newblock \showarticletitle{{Influence of soundtrack on eye movements during
  video exploration}}.
\newblock \bibinfo{journal}{{\em {Journal of Eye Movement Research}\/}}
  \bibinfo{volume}{5}, \bibinfo{number}{4} (\bibinfo{date}{Aug.}
  \bibinfo{year}{2012}), \bibinfo{pages}{2}.
\newblock
\showURL{%
\url{https://hal.archives-ouvertes.fr/hal-00723883}}
\newblock
\shownote{10 pages.}


\bibitem[\protect\citeauthoryear{Cummings and Bailenson}{Cummings and
  Bailenson}{2016}]%
        {cummings2016immersive}
\bibfield{author}{\bibinfo{person}{James~J Cummings} {and}
  \bibinfo{person}{Jeremy~N Bailenson}.} \bibinfo{year}{2016}\natexlab{}.
\newblock \showarticletitle{How immersive is enough? A meta-analysis of the
  effect of immersive technology on user presence}.
\newblock \bibinfo{journal}{{\em Media Psychology\/}} \bibinfo{volume}{19},
  \bibinfo{number}{2} (\bibinfo{year}{2016}), \bibinfo{pages}{272--309}.
\newblock


\bibitem[\protect\citeauthoryear{Damen, Leelasawassuk, Haines, Calway, and
  Mayol-Cuevas}{Damen et~al\mbox{.}}{2014}]%
        {damen2014you}
\bibfield{author}{\bibinfo{person}{Dima Damen}, \bibinfo{person}{Teesid
  Leelasawassuk}, \bibinfo{person}{Osian Haines}, \bibinfo{person}{Andrew
  Calway}, {and} \bibinfo{person}{Walterio Mayol-Cuevas}.}
  \bibinfo{year}{2014}\natexlab{}.
\newblock \showarticletitle{You-Do, I-Learn: Discovering Task Relevant Objects
  and their Modes of Interaction from Multi-User Egocentric Video}. In
  \bibinfo{booktitle}{{\em Proceedings of the British Machine Vision
  Conference}}. \bibinfo{publisher}{BMVA Press}.
\newblock
\showDOI{%
\url{https://doi.org/10.5244/C.28.30}}


\bibitem[\protect\citeauthoryear{David, Guti{\'e}rrez, Coutrot, Da~Silva, and
  Callet}{David et~al\mbox{.}}{2018}]%
        {david2018dataset}
\bibfield{author}{\bibinfo{person}{Erwan~J David}, \bibinfo{person}{Jes{\'u}s
  Guti{\'e}rrez}, \bibinfo{person}{Antoine Coutrot},
  \bibinfo{person}{Matthieu~Perreira Da~Silva}, {and}
  \bibinfo{person}{Patrick~Le Callet}.} \bibinfo{year}{2018}\natexlab{}.
\newblock \showarticletitle{A dataset of head and eye movements for {\ang{360}}
  videos}. In \bibinfo{booktitle}{{\em Proceedings of the 9th ACM Multimedia
  Systems Conference}}. ACM, \bibinfo{pages}{432--437}.
\newblock


\bibitem[\protect\citeauthoryear{Dorr, Martinetz, Gegenfurtner, and Barth}{Dorr
  et~al\mbox{.}}{2010}]%
        {dorr2010variability}
\bibfield{author}{\bibinfo{person}{Michael Dorr}, \bibinfo{person}{Thomas
  Martinetz}, \bibinfo{person}{Karl~R Gegenfurtner}, {and}
  \bibinfo{person}{Erhardt Barth}.} \bibinfo{year}{2010}\natexlab{}.
\newblock \showarticletitle{Variability of eye movements when viewing dynamic
  natural scenes}.
\newblock \bibinfo{journal}{{\em Journal of Vision\/}} \bibinfo{volume}{10},
  \bibinfo{number}{10} (\bibinfo{year}{2010}), \bibinfo{pages}{28--28}.
\newblock


\bibitem[\protect\citeauthoryear{Fang, Nakashima, Matsumiya, Kuriki, and
  Shioiri}{Fang et~al\mbox{.}}{2015}]%
        {fang2015eye}
\bibfield{author}{\bibinfo{person}{Yu Fang}, \bibinfo{person}{Ryoichi
  Nakashima}, \bibinfo{person}{Kazumichi Matsumiya}, \bibinfo{person}{Ichiro
  Kuriki}, {and} \bibinfo{person}{Satoshi Shioiri}.}
  \bibinfo{year}{2015}\natexlab{}.
\newblock \showarticletitle{Eye-head coordination for visual cognitive
  processing}.
\newblock \bibinfo{journal}{{\em PLOS ONE\/}} \bibinfo{volume}{10},
  \bibinfo{number}{3} (\bibinfo{year}{2015}), \bibinfo{pages}{e0121035}.
\newblock


\bibitem[\protect\citeauthoryear{Ferman, Collewijn, Jansen, and den
  Berg}{Ferman et~al\mbox{.}}{1987}]%
        {ferman1987human}
\bibfield{author}{\bibinfo{person}{L. Ferman}, \bibinfo{person}{H. Collewijn},
  \bibinfo{person}{T.C. Jansen}, {and} \bibinfo{person}{A.V.~Van den Berg}.}
  \bibinfo{year}{1987}\natexlab{}.
\newblock \showarticletitle{Human gaze stability in the horizontal, vertical
  and torsional direction during voluntary head movements, evaluated with a
  three-dimensional scleral induction coil technique}.
\newblock \bibinfo{journal}{{\em Vision Research\/}} \bibinfo{volume}{27},
  \bibinfo{number}{5} (\bibinfo{year}{1987}), \bibinfo{pages}{811 -- 828}.
\newblock
\showISSN{0042-6989}
\showDOI{%
\url{https://doi.org/10.1016/0042-6989(87)90078-2}}


\bibitem[\protect\citeauthoryear{Fischer, Jin~Chang, and Demiris}{Fischer
  et~al\mbox{.}}{2018}]%
        {fischer2018rtgene}
\bibfield{author}{\bibinfo{person}{Tobias Fischer}, \bibinfo{person}{Hyung
  Jin~Chang}, {and} \bibinfo{person}{Yiannis Demiris}.}
  \bibinfo{year}{2018}\natexlab{}.
\newblock \showarticletitle{RT-GENE: Real-Time Eye Gaze Estimation in Natural
  Environments}. In \bibinfo{booktitle}{{\em The European Conference on
  Computer Vision (ECCV)}}.
\newblock


\bibitem[\protect\citeauthoryear{Guti\'errez, David, Rai, and
  Callet}{Guti\'errez et~al\mbox{.}}{2018}]%
        {gutierrez2018toolbox}
\bibfield{author}{\bibinfo{person}{Jes\'us Guti\'errez}, \bibinfo{person}{Erwan
  David}, \bibinfo{person}{Yashas Rai}, {and} \bibinfo{person}{Patrick~Le
  Callet}.} \bibinfo{year}{2018}\natexlab{}.
\newblock \showarticletitle{Toolbox and dataset for the development of saliency
  and scanpath models for omnidirectional/{\ang{360}} still images}.
\newblock \bibinfo{journal}{{\em Signal Processing: Image Communication\/}}
  \bibinfo{volume}{69} (\bibinfo{year}{2018}), \bibinfo{pages}{35 -- 42}.
\newblock
\showISSN{0923-5965}
\showDOI{%
\url{https://doi.org/10.1016/j.image.2018.05.003}}
\newblock
\shownote{Salient360: Visual attention modeling for {\ang{360}} Images.}


\bibitem[\protect\citeauthoryear{Hessels, Niehorster, Nystr{\"o}m, Andersson,
  and Hooge}{Hessels et~al\mbox{.}}{2018}]%
        {hessels2018eye}
\bibfield{author}{\bibinfo{person}{Roy~S. Hessels},
  \bibinfo{person}{Diederick~C. Niehorster}, \bibinfo{person}{Marcus
  Nystr{\"o}m}, \bibinfo{person}{Richard Andersson}, {and}
  \bibinfo{person}{Ignace T.~C. Hooge}.} \bibinfo{year}{2018}\natexlab{}.
\newblock \showarticletitle{Is the eye-movement field confused about fixations
  and saccades? A survey among 124 researchers}.
\newblock \bibinfo{journal}{{\em Royal Society Open Science\/}}
  \bibinfo{volume}{5}, \bibinfo{number}{8} (\bibinfo{year}{2018}).
\newblock
\showDOI{%
\url{https://doi.org/10.1098/rsos.180502}}
\showeprint{http://rsos.royalsocietypublishing.org/content/5/8/180502.full.pdf}


\bibitem[\protect\citeauthoryear{Hooge, Niehorster, Nystr{\"o}m, Andersson, and
  Hessels}{Hooge et~al\mbox{.}}{2017}]%
        {hooge2017human}
\bibfield{author}{\bibinfo{person}{Ignace T.~C. Hooge},
  \bibinfo{person}{Diederick~C. Niehorster}, \bibinfo{person}{Marcus
  Nystr{\"o}m}, \bibinfo{person}{Richard Andersson}, {and}
  \bibinfo{person}{Roy~S. Hessels}.} \bibinfo{year}{2017}\natexlab{}.
\newblock \showarticletitle{Is human classification by experienced untrained
  observers a gold standard in fixation detection?}
\newblock \bibinfo{journal}{{\em Behavior Research Methods\/}}
  (\bibinfo{date}{19 Oct} \bibinfo{year}{2017}).
\newblock
\showISSN{1554-3528}
\showDOI{%
\url{https://doi.org/10.3758/s13428-017-0955-x}}


\bibitem[\protect\citeauthoryear{Hossain and Mil{\'e}us}{Hossain and
  Mil{\'e}us}{2016}]%
        {hossain2016eye}
\bibfield{author}{\bibinfo{person}{Akdas Hossain} {and} \bibinfo{person}{Emma
  Mil{\'e}us}.} \bibinfo{year}{2016}\natexlab{}.
\newblock \bibinfo{title}{Eye Movement Event Detection for Wearable Eye
  Trackers}.
\newblock   (\bibinfo{year}{2016}).
\newblock


\bibitem[\protect\citeauthoryear{Jennett, Cox, Cairns, Dhoparee, Epps, Tijs,
  and Walton}{Jennett et~al\mbox{.}}{2008}]%
        {jennett2008measuring}
\bibfield{author}{\bibinfo{person}{Charlene Jennett}, \bibinfo{person}{Anna~L
  Cox}, \bibinfo{person}{Paul Cairns}, \bibinfo{person}{Samira Dhoparee},
  \bibinfo{person}{Andrew Epps}, \bibinfo{person}{Tim Tijs}, {and}
  \bibinfo{person}{Alison Walton}.} \bibinfo{year}{2008}\natexlab{}.
\newblock \showarticletitle{Measuring and defining the experience of immersion
  in games}.
\newblock \bibinfo{journal}{{\em International Journal of Human-computer
  Studies\/}} \bibinfo{volume}{66}, \bibinfo{number}{9} (\bibinfo{year}{2008}),
  \bibinfo{pages}{641--661}.
\newblock


\bibitem[\protect\citeauthoryear{John}{John}{2017}]%
        {john2017dataset}
\bibfield{author}{\bibinfo{person}{Brendan John}.}
  \bibinfo{year}{2017}\natexlab{}.
\newblock \bibinfo{title}{A Dataset of Gaze Behavior in VR Faithful to Natural
  Statistics}.
\newblock \bibinfo{howpublished}{Rochester Institute of Technology}.
  (\bibinfo{year}{2017}).
\newblock


\bibitem[\protect\citeauthoryear{Kinsman, Evans, Sweeney, Keane, and
  Pelz}{Kinsman et~al\mbox{.}}{2012}]%
        {kinsman2012ego}
\bibfield{author}{\bibinfo{person}{Thomas Kinsman}, \bibinfo{person}{Karen
  Evans}, \bibinfo{person}{Glenn Sweeney}, \bibinfo{person}{Tommy Keane}, {and}
  \bibinfo{person}{Jeff Pelz}.} \bibinfo{year}{2012}\natexlab{}.
\newblock \showarticletitle{Ego-motion compensation improves fixation detection
  in wearable eye tracking}. In \bibinfo{booktitle}{{\em Proceedings of the
  Symposium on Eye Tracking Research and Applications}}. ACM,
  \bibinfo{pages}{221--224}.
\newblock


\bibitem[\protect\citeauthoryear{Kothari, Binaee, Bailey, Kanan, Diaz, and
  Pelz}{Kothari et~al\mbox{.}}{2017}]%
        {kothari2017gaze}
\bibfield{author}{\bibinfo{person}{Rakshit Kothari}, \bibinfo{person}{Kamran
  Binaee}, \bibinfo{person}{Reynold Bailey}, \bibinfo{person}{Christopher
  Kanan}, \bibinfo{person}{Gabriel Diaz}, {and} \bibinfo{person}{Jeff Pelz}.}
  \bibinfo{year}{2017}\natexlab{}.
\newblock \showarticletitle{Gaze-in-World movement Classification for
  Unconstrained Head Motion during Natural Tasks}.
\newblock \bibinfo{journal}{{\em Journal of Vision\/}}  \bibinfo{volume}{17}
  (\bibinfo{date}{08} \bibinfo{year}{2017}), \bibinfo{pages}{1156}.
\newblock
\showDOI{%
\url{https://doi.org/10.1167/17.10.1156}}


\bibitem[\protect\citeauthoryear{Larsson, Nystr\"om, Ard\"o, \r{A}str\"om, and
  Stridh}{Larsson et~al\mbox{.}}{2016}]%
        {larsson2016smooth}
\bibfield{author}{\bibinfo{person}{Linn\'ea Larsson}, \bibinfo{person}{Marcus
  Nystr\"om}, \bibinfo{person}{H\r{a}kan Ard\"o}, \bibinfo{person}{Kalle
  \r{A}str\"om}, {and} \bibinfo{person}{Martin Stridh}.}
  \bibinfo{year}{2016}\natexlab{}.
\newblock \showarticletitle{Smooth pursuit detection in binocular eye-tracking
  data with automatic video-based performance evaluation}.
\newblock \bibinfo{journal}{{\em Journal of Vision\/}} \bibinfo{volume}{16},
  \bibinfo{number}{15} (\bibinfo{year}{2016}), \bibinfo{pages}{20}.
\newblock
\showDOI{%
\url{https://doi.org/10.1167/16.15.20}}


\bibitem[\protect\citeauthoryear{Larsson, Nystr{\"o}m, and Stridh}{Larsson
  et~al\mbox{.}}{2013}]%
        {larsson2013detection}
\bibfield{author}{\bibinfo{person}{Linn\'ea Larsson}, \bibinfo{person}{Marcus
  Nystr{\"o}m}, {and} \bibinfo{person}{Martin Stridh}.}
  \bibinfo{year}{2013}\natexlab{}.
\newblock \showarticletitle{Detection of Saccades and Postsaccadic Oscillations
  in the Presence of Smooth Pursuit}.
\newblock \bibinfo{journal}{{\em IEEE Transactions on Biomedical
  Engineering\/}} \bibinfo{volume}{60}, \bibinfo{number}{9}
  (\bibinfo{date}{Sept} \bibinfo{year}{2013}), \bibinfo{pages}{2484--2493}.
\newblock
\showISSN{0018-9294}
\showDOI{%
\url{https://doi.org/10.1109/TBME.2013.2258918}}


\bibitem[\protect\citeauthoryear{Lee, Ghosh, and Grauman}{Lee
  et~al\mbox{.}}{2012}]%
        {lee2012discovering}
\bibfield{author}{\bibinfo{person}{Y.~J. Lee}, \bibinfo{person}{J. Ghosh},
  {and} \bibinfo{person}{K. Grauman}.} \bibinfo{year}{2012}\natexlab{}.
\newblock \showarticletitle{Discovering important people and objects for
  egocentric video summarization}. In \bibinfo{booktitle}{{\em 2012 IEEE
  Conference on Computer Vision and Pattern Recognition}}.
  \bibinfo{pages}{1346--1353}.
\newblock
\showISSN{1063-6919}
\showDOI{%
\url{https://doi.org/10.1109/CVPR.2012.6247820}}


\bibitem[\protect\citeauthoryear{Li, Kanemura, Asoh, Miyanishi, and
  Kawanabe}{Li et~al\mbox{.}}{2018}]%
        {li2018sparse}
\bibfield{author}{\bibinfo{person}{Y. Li}, \bibinfo{person}{A. Kanemura},
  \bibinfo{person}{H. Asoh}, \bibinfo{person}{T. Miyanishi}, {and}
  \bibinfo{person}{M. Kawanabe}.} \bibinfo{year}{2018}\natexlab{}.
\newblock \showarticletitle{A Sparse Coding Framework for Gaze Prediction in
  Egocentric Video}. In \bibinfo{booktitle}{{\em 2018 IEEE International
  Conference on Acoustics, Speech and Signal Processing (ICASSP)}}.
  \bibinfo{pages}{1313--1317}.
\newblock
\showISSN{2379-190X}
\showDOI{%
\url{https://doi.org/10.1109/ICASSP.2018.8462640}}


\bibitem[\protect\citeauthoryear{Lo, Fan, Lee, Huang, Chen, and Hsu}{Lo
  et~al\mbox{.}}{2017}]%
        {lo2017video}
\bibfield{author}{\bibinfo{person}{Wen-Chih Lo}, \bibinfo{person}{Ching-Ling
  Fan}, \bibinfo{person}{Jean Lee}, \bibinfo{person}{Chun-Ying Huang},
  \bibinfo{person}{Kuan-Ta Chen}, {and} \bibinfo{person}{Cheng-Hsin Hsu}.}
  \bibinfo{year}{2017}\natexlab{}.
\newblock \showarticletitle{{\ang{360}} Video Viewing Dataset in Head-Mounted
  Virtual Reality}. In \bibinfo{booktitle}{{\em Proceedings of the 8th ACM on
  Multimedia Systems Conference}} {\em (\bibinfo{series}{MMSys'17})}.
  \bibinfo{publisher}{ACM}, \bibinfo{address}{New York, NY, USA},
  \bibinfo{pages}{211--216}.
\newblock
\showISBNx{978-1-4503-5002-0}
\showDOI{%
\url{https://doi.org/10.1145/3083187.3083219}}


\bibitem[\protect\citeauthoryear{L{\"o}we, Stengel, F{\"o}rster, Grogorick, and
  Magnor}{L{\"o}we et~al\mbox{.}}{2015}]%
        {lowe2015visualization}
\bibfield{author}{\bibinfo{person}{Thomas L{\"o}we}, \bibinfo{person}{Michael
  Stengel}, \bibinfo{person}{Emmy-Charlotte F{\"o}rster},
  \bibinfo{person}{Steve Grogorick}, {and} \bibinfo{person}{Marcus Magnor}.}
  \bibinfo{year}{2015}\natexlab{}.
\newblock \showarticletitle{Visualization and analysis of head movement and
  gaze data for immersive video in head-mounted displays}. In
  \bibinfo{booktitle}{{\em Proceedings of the Workshop on Eye Tracking and
  Visualization (ETVIS)}}.
\newblock


\bibitem[\protect\citeauthoryear{Meyer, Lasker, and Robinson}{Meyer
  et~al\mbox{.}}{1985}]%
        {meyer1985upper}
\bibfield{author}{\bibinfo{person}{Craig~H. Meyer}, \bibinfo{person}{Adrian~G.
  Lasker}, {and} \bibinfo{person}{David~A. Robinson}.}
  \bibinfo{year}{1985}\natexlab{}.
\newblock \showarticletitle{The upper limit of human smooth pursuit velocity}.
\newblock \bibinfo{journal}{{\em Vision Research\/}} \bibinfo{volume}{25},
  \bibinfo{number}{4} (\bibinfo{year}{1985}), \bibinfo{pages}{561 -- 563}.
\newblock
\showISSN{0042-6989}
\showDOI{%
\url{https://doi.org/10.1016/0042-6989(85)90160-9}}


\bibitem[\protect\citeauthoryear{Nguyen, Yan, and Nahrstedt}{Nguyen
  et~al\mbox{.}}{2018}]%
        {nguyen2018attention}
\bibfield{author}{\bibinfo{person}{Anh Nguyen}, \bibinfo{person}{Zhisheng Yan},
  {and} \bibinfo{person}{Klara Nahrstedt}.} \bibinfo{year}{2018}\natexlab{}.
\newblock \showarticletitle{Your Attention is Unique: Detecting 360-Degree
  Video Saliency in Head-Mounted Display for Head Movement Prediction}. In
  \bibinfo{booktitle}{{\em Proceedings of the 26th ACM International Conference
  on Multimedia}} {\em (\bibinfo{series}{MM '18})}. \bibinfo{publisher}{ACM},
  \bibinfo{address}{New York, NY, USA}, \bibinfo{pages}{1190--1198}.
\newblock
\showISBNx{978-1-4503-5665-7}
\showDOI{%
\url{https://doi.org/10.1145/3240508.3240669}}


\bibitem[\protect\citeauthoryear{Olsen}{Olsen}{2012}]%
        {olsen2012tobii}
\bibfield{author}{\bibinfo{person}{Anneli Olsen}.}
  \bibinfo{year}{2012}\natexlab{}.
\newblock \showarticletitle{The Tobii I-VT fixation filter}.
\newblock \bibinfo{journal}{{\em Tobii Technology\/}} (\bibinfo{year}{2012}).
\newblock


\bibitem[\protect\citeauthoryear{Polatsek, Benesova, Paletta, and
  Perko}{Polatsek et~al\mbox{.}}{2016}]%
        {polatsek2016novelty}
\bibfield{author}{\bibinfo{person}{P. Polatsek}, \bibinfo{person}{W. Benesova},
  \bibinfo{person}{L. Paletta}, {and} \bibinfo{person}{R. Perko}.}
  \bibinfo{year}{2016}\natexlab{}.
\newblock \showarticletitle{Novelty-based Spatiotemporal Saliency Detection for
  Prediction of Gaze in Egocentric Video}.
\newblock \bibinfo{journal}{{\em IEEE Signal Processing Letters\/}}
  \bibinfo{volume}{23}, \bibinfo{number}{3} (\bibinfo{date}{March}
  \bibinfo{year}{2016}), \bibinfo{pages}{394--398}.
\newblock
\showISSN{1070-9908}
\showDOI{%
\url{https://doi.org/10.1109/LSP.2016.2523339}}


\bibitem[\protect\citeauthoryear{Rai, Guti{\'e}rrez, and Le~Callet}{Rai
  et~al\mbox{.}}{2017}]%
        {rai2017dataset}
\bibfield{author}{\bibinfo{person}{Yashas Rai}, \bibinfo{person}{Jes{\'u}s
  Guti{\'e}rrez}, {and} \bibinfo{person}{Patrick Le~Callet}.}
  \bibinfo{year}{2017}\natexlab{}.
\newblock \showarticletitle{A dataset of head and eye movements for 360 degree
  images}. In \bibinfo{booktitle}{{\em Proceedings of the 8th ACM on Multimedia
  Systems Conference}}. ACM, \bibinfo{pages}{205--210}.
\newblock


\bibitem[\protect\citeauthoryear{Salvucci and Goldberg}{Salvucci and
  Goldberg}{2000}]%
        {salvucci2000identifying}
\bibfield{author}{\bibinfo{person}{Dario~D. Salvucci} {and}
  \bibinfo{person}{Joseph~H. Goldberg}.} \bibinfo{year}{2000}\natexlab{}.
\newblock \showarticletitle{Identifying Fixations and Saccades in Eye-tracking
  Protocols}. In \bibinfo{booktitle}{{\em Proceedings of the 2000 Symposium on
  Eye Tracking Research \& Applications}} {\em (\bibinfo{series}{ETRA '00})}.
  \bibinfo{publisher}{ACM}, \bibinfo{address}{New York, NY, USA},
  \bibinfo{pages}{71--78}.
\newblock
\showISBNx{1-58113-280-8}
\showDOI{%
\url{https://doi.org/10.1145/355017.355028}}


\bibitem[\protect\citeauthoryear{Santini, Fuhl, K\"{u}bler, and
  Kasneci}{Santini et~al\mbox{.}}{2016}]%
        {santini2016bayesian}
\bibfield{author}{\bibinfo{person}{Thiago Santini}, \bibinfo{person}{Wolfgang
  Fuhl}, \bibinfo{person}{Thomas K\"{u}bler}, {and} \bibinfo{person}{Enkelejda
  Kasneci}.} \bibinfo{year}{2016}\natexlab{}.
\newblock \showarticletitle{Bayesian Identification of Fixations, Saccades, and
  Smooth Pursuits}. In \bibinfo{booktitle}{{\em {Proceedings of the Ninth
  Biennial ACM Symposium on Eye Tracking Research \& Applications}}} {\em
  (\bibinfo{series}{ETRA '16})}. \bibinfo{publisher}{ACM},
  \bibinfo{address}{New York, NY, USA}, \bibinfo{pages}{163--170}.
\newblock
\showISBNx{978-1-4503-4125-7}
\showDOI{%
\url{https://doi.org/10.1145/2857491.2857512}}


\bibitem[\protect\citeauthoryear{Sitzmann, Serrano, Pavel, Agrawala, Gutierrez,
  Masia, and Wetzstein}{Sitzmann et~al\mbox{.}}{2018}]%
        {sitzmann2018saliency}
\bibfield{author}{\bibinfo{person}{V. Sitzmann}, \bibinfo{person}{A. Serrano},
  \bibinfo{person}{A. Pavel}, \bibinfo{person}{M. Agrawala},
  \bibinfo{person}{D. Gutierrez}, \bibinfo{person}{B. Masia}, {and}
  \bibinfo{person}{G. Wetzstein}.} \bibinfo{year}{2018}\natexlab{}.
\newblock \showarticletitle{Saliency in VR: How Do People Explore Virtual
  Environments?}
\newblock \bibinfo{journal}{{\em IEEE Transactions on Visualization and
  Computer Graphics\/}} \bibinfo{volume}{24}, \bibinfo{number}{4}
  (\bibinfo{date}{April} \bibinfo{year}{2018}), \bibinfo{pages}{1633--1642}.
\newblock
\showISSN{1077-2626}
\showDOI{%
\url{https://doi.org/10.1109/TVCG.2018.2793599}}


\bibitem[\protect\citeauthoryear{Startsev, Agtzidis, and Dorr}{Startsev
  et~al\mbox{.}}{2018}]%
        {startsev2018cnn}
\bibfield{author}{\bibinfo{person}{Mikhail Startsev}, \bibinfo{person}{Ioannis
  Agtzidis}, {and} \bibinfo{person}{Michael Dorr}.}
  \bibinfo{year}{2018}\natexlab{}.
\newblock \showarticletitle{1D CNN with BLSTM for automated classification of
  fixations, saccades, and smooth pursuits}.
\newblock \bibinfo{journal}{{\em Behavior Research Methods\/}}
  (\bibinfo{date}{08 Nov} \bibinfo{year}{2018}).
\newblock
\showISSN{1554-3528}
\showDOI{%
\url{https://doi.org/10.3758/s13428-018-1144-2}}


\bibitem[\protect\citeauthoryear{Steil, Huang, and Bulling}{Steil
  et~al\mbox{.}}{2018}]%
        {steil2018fixation}
\bibfield{author}{\bibinfo{person}{Julian Steil},
  \bibinfo{person}{Michael~Xuelin Huang}, {and} \bibinfo{person}{Andreas
  Bulling}.} \bibinfo{year}{2018}\natexlab{}.
\newblock \showarticletitle{Fixation Detection for Head-mounted Eye Tracking
  Based on Visual Similarity of Gaze Targets}. In \bibinfo{booktitle}{{\em
  Proceedings of the 2018 ACM Symposium on Eye Tracking Research \&
  Applications}} {\em (\bibinfo{series}{ETRA '18})}. \bibinfo{publisher}{ACM},
  \bibinfo{address}{New York, NY, USA}, \bibinfo{pages}{23:1--23:9}.
\newblock
\showISBNx{978-1-4503-5706-7}
\showDOI{%
\url{https://doi.org/10.1145/3204493.3204538}}


\bibitem[\protect\citeauthoryear{Thaler, Sch\"utz, Goodale, and
  Gegenfurtner}{Thaler et~al\mbox{.}}{2013}]%
        {thaler2013best}
\bibfield{author}{\bibinfo{person}{L. Thaler}, \bibinfo{person}{A.C. Sch\"utz},
  \bibinfo{person}{M.A. Goodale}, {and} \bibinfo{person}{K.R. Gegenfurtner}.}
  \bibinfo{year}{2013}\natexlab{}.
\newblock \showarticletitle{What is the best fixation target? The effect of
  target shape on stability of fixational eye movements}.
\newblock \bibinfo{journal}{{\em Vision Research\/}}  \bibinfo{volume}{76}
  (\bibinfo{year}{2013}), \bibinfo{pages}{31--42}.
\newblock
\showISSN{0042-6989}
\showDOI{%
\url{https://doi.org/10.1016/j.visres.2012.10.012}}


\bibitem[\protect\citeauthoryear{Turuwhenua, Yu, Mazharullah, and
  Thompson}{Turuwhenua et~al\mbox{.}}{2014}]%
        {turu2014method}
\bibfield{author}{\bibinfo{person}{Jason Turuwhenua}, \bibinfo{person}{Tzu-Ying
  Yu}, \bibinfo{person}{Zan Mazharullah}, {and} \bibinfo{person}{Benjamin
  Thompson}.} \bibinfo{year}{2014}\natexlab{}.
\newblock \showarticletitle{A method for detecting optokinetic nystagmus based
  on the optic flow of the limbus}.
\newblock \bibinfo{journal}{{\em Vision Research\/}}  \bibinfo{volume}{103}
  (\bibinfo{year}{2014}), \bibinfo{pages}{75--82}.
\newblock


\bibitem[\protect\citeauthoryear{Zagoruyko and Komodakis}{Zagoruyko and
  Komodakis}{2015}]%
        {zagoruyko2015learning}
\bibfield{author}{\bibinfo{person}{Sergey Zagoruyko} {and}
  \bibinfo{person}{Nikos Komodakis}.} \bibinfo{year}{2015}\natexlab{}.
\newblock \showarticletitle{Learning to Compare Image Patches via Convolutional
  Neural Networks}. In \bibinfo{booktitle}{{\em The IEEE Conference on Computer
  Vision and Pattern Recognition (CVPR)}}.
\newblock


\end{thebibliography}

\end{document}